\documentclass[aps,prb,preprint,showpacs,preprintnumbers,amsmath,amssymb]{revtex4}
\usepackage{graphicx}
\begin{document}

\title{Rigorous formulation of  oblique incidence scattering from  dispersive media}

\author{Lingxiao Zhang and Tamar Seideman\footnote{Author to whom correspondence may be addressed, t-seideman@northwestern.edu}}
\affiliation{Departments of Chemistry and Physics,
Northwestern University, Evanston, Illinois 60208}

\date{\today}

\begin{abstract}

We formulate a  finite-difference
time-domain (FDTD) approach to simulate electromagnetic wave scattering from
scatterers embedded in layered dielectric or dispersive
media. At the heart of our approach is a derivation of an equivalent one-dimensional
wave propagation equation for dispersive media characterized by a linear
sum of Debye-, Drude- and Lorentz-type  poles. The derivation is followed by a detailed discussion
of the  simulation setup and numerical issues. The developed methodology
is tested  by comparison with analytical reflection and
transmission coefficients for scattering from a slab, illustrating good convergence behavior. The case of scattering from a sub-wavelength
slit in a dispersive thin film is explored to demonstrate the
applicability of our formulation to time- and incident angle-dependent
analysis of surface  waves generated by an obliquely incident
plane wave.

\end{abstract}


\maketitle

\section{INTRODUCTION}
\label{sec:sec1}

The study of obliquely incident plane wave upon planar interfaces is
of fundemantal interest to electromagnetic (EM) wave propagation. It
underlies Snell's law of refraction and leads to important
concepts such as total reflection
and Brewster's angle.\cite{Jackson, Wolf} One can easily
relate to the concept of an obliquely
incident plane wave by the daily experience
of  looking into a mirror. In practice, oblique incidence
is vastly applied in EM related applications such as fiber optics,\cite{Hecht}
underground object detection,\cite{Jol}
and RF-human body interaction.\cite{Vorst}

In the growing field of nanoplasmonics,\cite{Barnes, Ozbay} oblique incidence
finds applications particularly in exciting surface plasmon polaritons (SPPs),
exemplified by the common experimental setup in which subwavelength defects
or attenuated total reflection are utilized to couple the obliquely incident
plane wave into propagating
SPPs.\cite{Lopez,Coe}
By taking advantage of the incident angle degree of freedom,
several experiments have demonstrated SPP near-field
manipulation,\cite{Yin,Liu,Doui} which has been proposed as
a direct approach to measuring SPP generation efficiency.\cite{Wang}
Most recently, it has been shown that SPP's can be directly
generated on a planar metal surface
by interfering incoming light beams with different
incident angles in a four-wave mixing scheme.\cite{Renger}

The
effects of light incident at oblique angle  on sub-wavelength defects
in metallic layered media have been studied by
frequency-domain calculations based on either coupled wave analysis or a
semi-analytical model. These references have explored the
obliquely incident light
transmission through a single defect,\cite{Vidal1,Vidal2,Gordon}
and the SPP generation efficiency.\cite{Lalanne,Kim}
Our work is largely motivated by recent experimental study of SPP dynamics
excited or controlled by a femto-second (fs) laser pulse
obliquely incident on a SPP propagating
interface.\cite{MacDonald, footnote0}
To describe such experiments,
a time-domain method is desirable because of the ultrafast nature
of the exciting or controlling laser pulse. The major challenge
in developing such a method,  is to accurately treat
the oblique incidence as well as the material dispersiveness.
This
poses a special challenge for mesh-based propagation method (such
as the finite-difference time domain method) because even
in the absence of the inhomogeneous media the wave
front not aligned with the Cartesian mesh is required to be
uniform and to have arbitrary incident angle and time profile.

In this paper, we develop a numerical method to rigorously
treat obliquely incident plane wave
scattering at embedded scatterers in layered
dielectric and dispersive media.
To the best of our knowledge, such a method
was not published as yet.\cite{ftnt1}
Targetted mainly at
time-domain studies of EM wave phenomena that involve SPP
excitation and propagation in metallic films, the developed
method is
formulated within the framework of the finite-difference
time-domain (FDTD) method. This method has enjoyed a wide range of   applications in the field
of nanoplasmonics,\cite{Taflove,Teixeira}
and its time-domain nature makes it particularly well suited to ultrafast phenomena.
Our treatment of the oblique plane wave
is an extension of the total field / scatter field (TF/SF) technique
to describe media characterized by the combination of
 Debye-, Drude- and Lorentz-type poles.\cite{Ung,Lee}
The TF/SF technique has been applied successfully in
the FDTD study of free-standing scatterers, layered dielectrics and dispersive
media describable by a single Debye pole.\cite{Taflove, Winton, Capoglu, Jiang}
It is based on the linearity
of Maxwell's equations and decomposes the total field into an incident and
a scattered field components,\cite{Taflove}
\begin{equation}
\psi^{\rm tot} = \psi^{\rm inc} + \psi^{\rm sca}.
\label{eqn:eqn0}
\end{equation}
\noindent By setting up an artificial boundary between the TF and SF
regions in the FDTD simulations, a plane wave of arbitrary time profile
and incident angle can be achieved by matching the known incident field
at the TF and SF boundary. In presenting our method,
we will focus on the derivation of
an equivalent one-dimensional (1D) wave equations for the TF/SF
boundary condition, suitable for various types of material
dispersiveness, and explain in detail the numerical considerations involved. This will be followed by extensive numerical tests of the convergence properties of the method. For clarity, several important concepts
from the previous literature are reemphasized.

This paper is organized as follows: In Section \ref{sec:sec2}, we derive
the equivalent 1D wave equations, show the numerical
flow chart for matching the TF/SF boundary
condition, and discuss several practical simulation details, including
stability, interface treatment, and leakage. Section
\ref{sec:sec3} tests our approach by comparison of numerical with analytical results for model problems. Finally,
concluding remarks are provided in Section \ref{sec:sec4}.

\section{THEORY AND NUMERICAL METHOD}
\label{sec:sec2}

In the following, we provide the equations and numerical method
for solving the transverse magnetic (TM) mode in two dimensions (magnetic field
perpendicular to the two-dimensional plane). Special emphasis
is placed on the TM mode because of its relevance to
SPP excitation.\cite{Raether}
The numerical approach for solving the transverse electric (TE) mode equations is similar
to that for the TM mode, and the corresponding equations
are given in Appendix \ref{Appx1}. The media considered are vacuum,
linear dielectric media (characterized by a dielectric constant), and
linear dispersive media (characterized by a finite sum of Debye, Lorentz,
and Drude types of poles).

\subsection{TM mode wave propagation: two-dimensional and equivalent
one-dimensional equations}
\label{sec2:subsec1}

Our starting point is Maxwell's equations in the frequency domain for the TM mode,
 \begin{gather}
 \frac{\partial E_{y}}{\partial x}-\frac{\partial E_{x}}{\partial y}
 =i\omega\mu_{0}H_{z},\label{eqn:eqn1}\\
 \frac{\partial H_{z}}{\partial y}=-i\omega\epsilon_0\epsilon(\omega)
 E_{x},\label{eqn:eqn2}\\
 \frac{\partial H_{z}}{\partial x}=i\omega\epsilon_0\epsilon(\omega)
 E_{y},\label{eqn:eqn3}
 \end{gather}
where the coordinate system is defined in
Fig. \ref{fig:f1}, $\epsilon_0$  is the free space permitivity, $\mu_0$ is the free space permeability, and
$\epsilon(\omega)$ is the dielectric function for a dispersive media, which reduces to a constant for vacuum and dielectric media.

In the case studies below, we assume a dispersive medium with a single
(non-zero) Drude pole $\epsilon_m=\epsilon(\infty)-\omega_D^2/(\omega^2+i\Gamma_D\omega)$
and provide two separate sets of equations for
solving Eqs. (\ref{eqn:eqn1}-\ref{eqn:eqn3}). The first set of equations is based on
the auxiliary differential equation (ADE) approach with
polarization currents to account for the dispersiveness. In this case,
we further assume that media other than vacuum are not extended into
the absorbing boundary, which allows us to use
Berenger's PML absorbing boundary condition.\cite{Berenger} The second set
of equations is formulated within the general context of the Uniaxial
Perfectly Matched Layers (UPML) absorbing boundary
conditions,\cite{Gedney} and involves
a different approach to treat the dispersiveness. In this case,
we can effectively absorb the outgoing waves exiting the simulation
domain in the dielectric and
dispersive media. Separate tests
have been done to ensure that the two approaches provide the same solution.\cite{footnote1} In the following, we
assume that  the 2D electric and magnetic fields propagate on the Yee
mesh with the dependence $u\vert^{n}_{i,j}=u(i\Delta x, j\Delta y, n\Delta t)$.
$\Delta x=\Delta y$ is the size of a unit cell, and $\Delta t$ is
unit time step. Details on the FDTD equations in both
ADE and UPML approaches are given in Appendix \ref{Appx2}.

If we now consider TM mode wave propagation with obliquely incident
plane wave on layered media with translational invariance, the 2D equations of motion can be
reduced to an equivalent 1D wave propagation problem along the
direction perpendicular to the interfaces between the
media.\cite{Winton, Capoglu} We proceed to
derive the equivalent 1D wave equation for the TM mode, which will serve as
a means of introducing incident fields along the TF/SF boundary. The corresponding derivation for the TE mode is provided in Appendix \ref{Appx1}.

Substituting Eq. (\ref{eqn:eqn3}) into Eq. (\ref{eqn:eqn1}) yields,
 \begin{equation}
 \frac{\partial E_{x}}{\partial y}=-i\omega\mu_0 H_z
 +\frac{1}{i\omega\epsilon_0\epsilon(\omega)}\frac{\partial^2 H_{z}}{\partial x^2}.
 \label{eqn:eqn4}
 \end{equation}
Because of the translational invariance and phase matching
across the interfaces between different layers,
${\partial^2 H_{z}}/{\partial x^2}=-k_x^2 H_z$, with
$k_x$ being a wavevector that is identical for waves in different
layers.\cite{footnote2} If we further assume
that an oblique plane wave is incident from a dielectric medium with
relative permitivity $\epsilon_{1r}$,
then $k_x=\omega\sqrt{\mu_0\epsilon_0\epsilon_{1r}}\sin(\theta)$, which can be substituted
into Eq. (\ref{eqn:eqn4}) to give,
 \begin{equation}
 \frac{\partial E_{x}}{\partial y}=-i\omega\mu_0
 \left[\frac{\epsilon(\omega)-\epsilon_{1r}\sin^2(\theta)}{\epsilon(\omega)}\right]H_z.
 \label{eqn:eqn5}
 \end{equation}

Equations (\ref{eqn:eqn2}) and (\ref{eqn:eqn5}) constitute a system of equations
for 1D TM wave propagation across the interfaces between the media. To translate
those equations into FDTD equations, Jiang et al. introduced a convenient
method to overcome the difficulty of time-domain convolution between the
term in the square bracket and $H_z$ in Eq. (\ref{eqn:eqn5}). In this method,\cite{Jiang}
Eq. (\ref{eqn:eqn5}) is first split into a pair of  equations as,
 \begin{gather}
 \frac{\partial E_{x}}{\partial y}=-i\omega\mu_0H_z^{\prime},\label{eqn:eqn6}\\
 H_z^{\prime}=\frac{\epsilon(\omega)-\epsilon_{1r}\sin^2(\theta)}{\epsilon(\omega)} H_z.
 \label{eqn:eqn7}
 \end{gather}
Equations (\ref{eqn:eqn2}), (\ref{eqn:eqn6}) and (\ref{eqn:eqn7}) then lead to the following set of FDTD equations,
\begin{eqnarray}
E_{x1D}\vert^{n+1}_{j}&=& a_{x1}E_{x1D}\vert^{n}_{j}
                      +a_{x2}(H_{z1D}\vert^{n+1/2}_{j+1/2}-H_{z1D}\vert^{n+1/2}_{j-1/2})\notag\\
                      &&+a_{x3}J_{x1D}\vert^{n+1/2}_{j},\label{eqn:eqn8}\\
J_{x1D}\vert^{n+3/2}_{j}&=&a_{x4}J_{x1D}\vert^{n+1/2}_{j}+a_{x5}E_{x1D}\vert^{n+1}_{j},\label{eqn:eqn9}\\
H^{\prime}_{z1D}\vert^{n+3/2}_{j+1/2}&=&b_{y1}H^{\prime}_{z1D}\vert^{n+1/2}_{j+1/2}
                                   +b_{y2}(E_{x1D}\vert^{n+1}_{j+1}-E_{x1D}\vert^{n+1}_{j}),\label{eqn:eqn10}\\
H_{z1D}\vert^{n+3/2}_{j+1/2}&=&b_{y3}H_{z1D}\vert^{n+1/2}_{j+1/2}
                                   +b_{y4}H_{z1D}\vert^{n-1/2}_{j+1/2}\notag\\
                                   &&+b_{y5}H^{\prime}_{z1D}\vert^{n+3/2}_{j+1/2}
                                   +b_{y6}H^{\prime}_{z1D}\vert^{n+1/2}_{j+1/2}\notag\\
                                   &&+b_{y7}H^{\prime}_{z1D}\vert^{n-1/2}_{j+1/2}.\label{eqn:eqn11}
\end{eqnarray}
In obtaining Eq. (\ref{eqn:eqn11}), we have multiplied both sides of Eq. (\ref{eqn:eqn7})
by $\epsilon(\omega)$ and Fourier transformed the result into the time domain. We have also made
the assumption that a Drude model is used,
$\epsilon_m=\epsilon(\infty)-\omega_D^2/(\omega^2+i\Gamma_D\omega)$. The updating
coefficients in Eq. (\ref{eqn:eqn11}) are
\begin{equation}
\begin{cases}
b_{y3}=b_{y4}=b_{y6}=b_{y7} = 0\\
b_{y5}=\epsilon_r/\left[\epsilon_r-\epsilon_{1r}\sin^2(\theta)\right]\label{eqn:eqn12}\\
\end{cases}
\end{equation}
in vacuum ($\epsilon_r=1$) and dielectric media (constant $\epsilon_r$), and
\begin{equation}
\begin{cases}
b_{y3}=\frac{4\left[\epsilon(\infty)-\epsilon_{1r}\sin^2(\theta)\right]}{(2
       +\Gamma_D\Delta t)\left[\epsilon(\infty)-\epsilon_{1r}\sin^2(\theta)\right]+\omega_D^2\Delta t^2},\\
b_{y4}=-\frac{(2-\Gamma_D\Delta t)\left[\epsilon(\infty)-\epsilon_{1r}\sin^2(\theta)\right]
              +\omega_D^2\Delta t^2}
             {(2+\Gamma_D\Delta t)\left[\epsilon(\infty)-\epsilon_{1r}\sin^2(\theta)\right]
             +\omega_D^2\Delta t^2},\\
b_{y5}=\frac{(2+\Gamma_D\Delta t)\epsilon(\infty)+\omega_D^2\Delta t^2}
            {(2+\Gamma_D\Delta t)\left[\epsilon(\infty)-\epsilon_{1r}\sin^2(\theta)\right]
             +\omega_D^2\Delta t^2},\\
b_{y6}=\frac{-4\epsilon(\infty)}
            {(2+\Gamma_D\Delta t)\left[\epsilon(\infty)-\epsilon_{1r}\sin^2(\theta)\right]
             +\omega_D^2\Delta t^2},\\
b_{y7}=\frac{(2-\Gamma_D\Delta t)\epsilon(\infty)+\omega_D^2\Delta t^2}
            {(2+\Gamma_D\Delta t)\left[\epsilon(\infty)-\epsilon_{1r}\sin^2(\theta)\right]
             +\omega_D^2\Delta t^2}\label{eqn:eqn13}\\
\end{cases}
\end{equation}
in Drude media.
Here, we note the similarity between the updates of the $(H^\prime,H)$ pair
and the $(P,D)$ pair in the UPML formulation, which results from the fact that
both pairs involves updating an auxiliary variable before the treatment
of the material dispersiveness.
In the case that $\epsilon(\omega)$
contains a linear sum of different types of poles (e.g., to accurately
describe metals near inter-band transition energies\cite{Lee}),
direct Fourier transform may not
be as efficient because of higher-order derivatives with respect to time.
For a systematic treatment of this situation, interested readers are referred
to Appendix \ref{Appx3}. The updating coefficients in Eqs. (\ref{eqn:eqn8})
to (\ref{eqn:eqn10}) are identical to those in Eqs. (\ref{eqn:a}), (\ref{eqn:b})
and (\ref{eqn:c}), which are given in Eqs. (\ref{eqn:d}-\ref{eqn:h}). We note that the updating
coefficients corresponding to Berenger's PML formulation can be used here provided
that the two end media in the layers are vacuum. In the case of
non-vacuum semi-infinite media at the two ends, 1D UPML can be used to effectively
absorb the outgoing waves, for example, equations similar to
Eqs. (\ref{eqn:l}-\ref{eqn:k}, \ref{eqn:i0}, \ref{eqn:i})
can be used by setting $\kappa_x=1$ and $\sigma_x=0$ in Eqs. (\ref{eqn:j1})
and (\ref{eqn:k2}).

\subsection{Simulation setup and flow chart}

\begin{figure}[htp]
\begin{center}
\includegraphics[width=11cm]{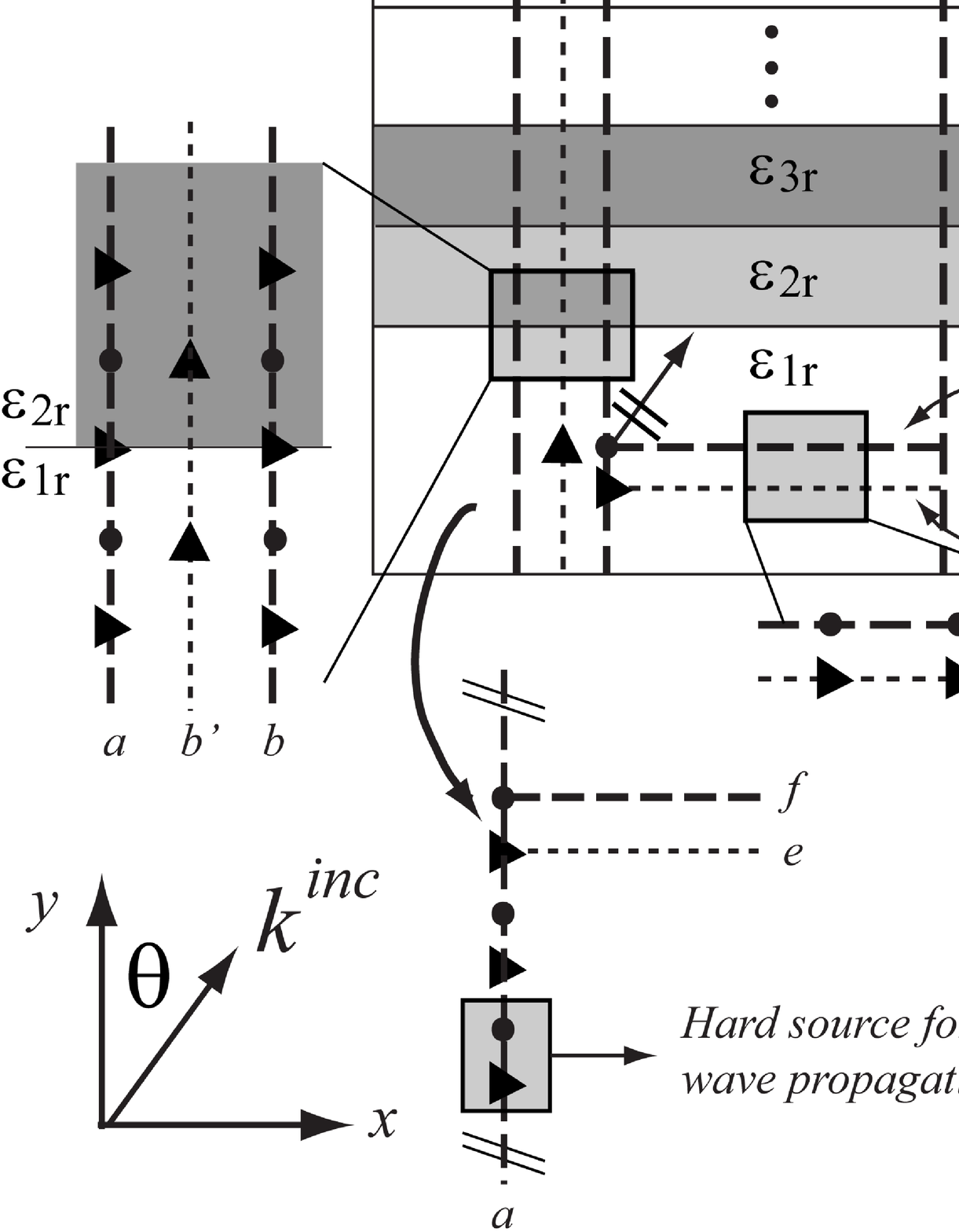}
\caption{Simulation geometry: the layered media are
distinguished by different shades and denoted by $\epsilon_{1r}$,
$\epsilon_{2r}$, etc. Thick, dashed (thin, dotted)
lines denote the boundaries to which the $H-$ ($E-$)
field is assigned. The left, lower left, and lower
right panels show the specific field point assignment
at the interface, along line $a$, and at the horizontal
boundaries, respectively.
The $x$-coordinates of lines $a$, $b^\prime$,
$b$, $c$, $c^\prime$, $d$ are $i_1-1/2$, $i_1$,
$i_1+1/2$, $i_2-1/2$, $i_2$, and $i_2+1/2$,
respectively. The $y$-coordinates of lines
$e$, $f$, $g$, and $h$ are $j_1$, $j_1+1/2$,
$j_2-1/2$, and $j_2$, respectively.}
\label{fig:f1}
\end{center}
\end{figure}

The main panel of Fig. \ref{fig:f1}
illustrates the geometry of the FDTD simulation region.
The layered media are denoted by $\epsilon_{1r}$, $\epsilon_{2r}$, etc.
and are stacked along the $y$ direction. The thick, dashed (thin, dotted) lines
denote the TF/SF boundaries along which the incident $H$-field ($E$-field) is
calculated. Incident field alignments on the boundaries
are shown more explicitly in the zoom-in panels to the left and below the
main panel. In this work, we assume that the oblique incidence field
is introduced from the lower left corner (crossing point
between lines $b$ and $f$ in the main panel of Fig. \ref{fig:f1})
with incident angle $\theta$ to the normal
of the media interfaces ($y$ direction). We further assume that the two end media in the
layers are vacuum. Consequently $\epsilon_{1r}=1$, so that
field propagation along the horizontal boundaries $e$ through $h$ can be
readily calculated by a delay of the free-space propagation time. In addition, the 1D field propagation along the vertical
lines can be terminated by Berenger's PML formulation. The perfectly matched layers
absorbing boundaries are not
shown in Fig. \ref{fig:f1}. They will be
further illustrated and explained when we consider specific examples in
Section \ref{sec:sec3}. The lower left panel in Fig. \ref{fig:f1} shows
the field alignment along line $a$ for the 1D wave propagation. The same
setup applies to lines $b$, $c$ and $d$. Importantly,
a 1D total field / scattered field approach is used here (the boundary points
are highlighted in the shaded rectangle) because we must allow the wave
from the multiple interface reflection to exit the 1D simulation and be
absorbed at the bottom on the 1D simulation line.\cite{Capoglu}

\begin{figure}[htp]
\begin{center}
\includegraphics[width=12cm]{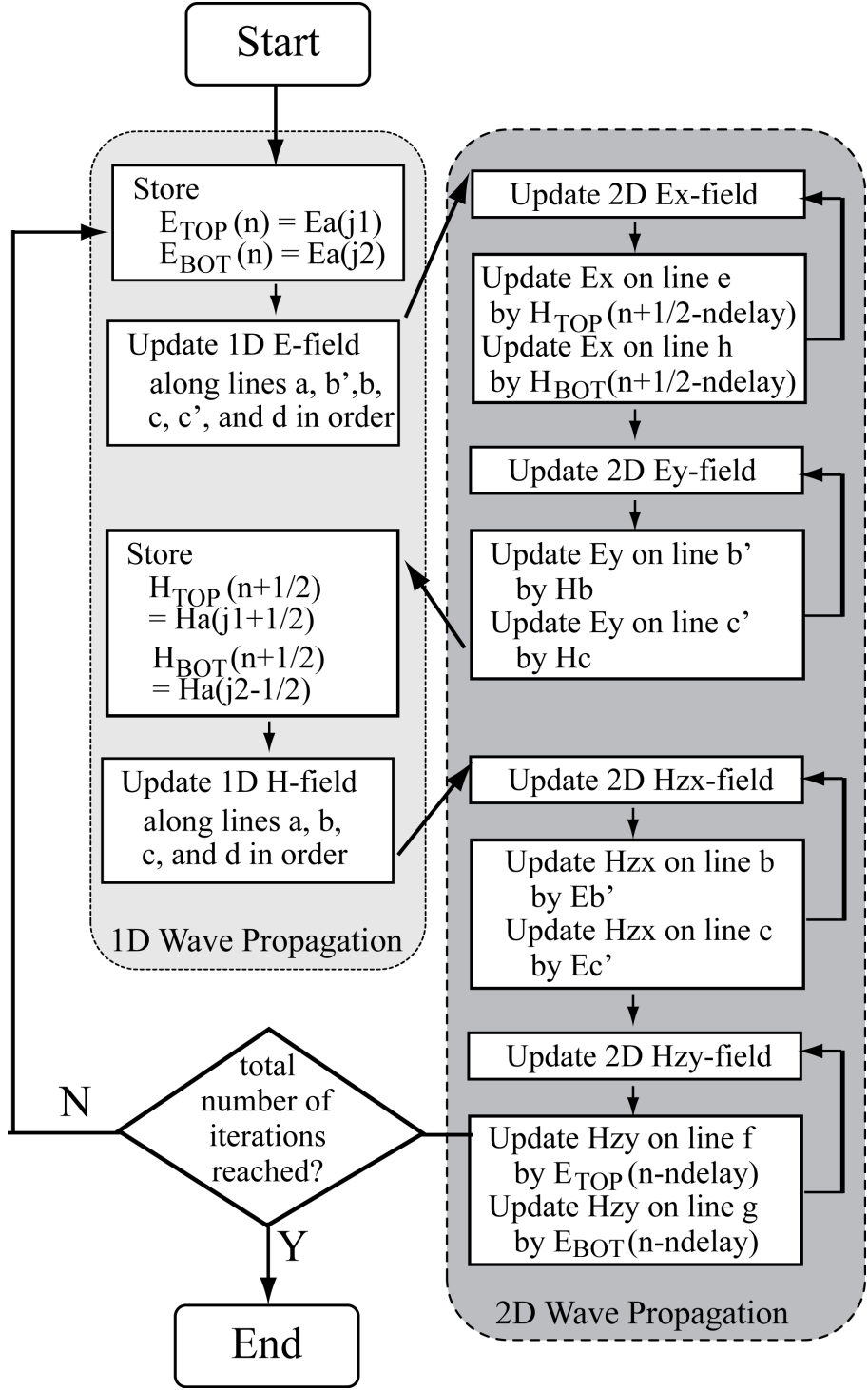}
\caption{Simulation flow chart. Note that ``TOP'' (``BOT'')
refers to a lower (higher) $y$ coordinate. For 1D field updates
an Auxiliary Differential Equation (ADE) approach is used, while 2D field
updates are performed by either the ADE approach or the equations
consistent with the
Uniaxial Perfectly Matched Layers (UPML) formulation.}
\label{fig:f2}
\end{center}
\end{figure}

Our simulation follows the flow chart shown in Fig. \ref{fig:f2}. The
procedures belonging to 1D and 2D field updates are highlighted in the
shaded rounded rectangles. In each iteration, the code updates the
1D $E$-field, 2D $E$-field, 1D $H$-field, and 2D $H$-field in a sequence.
The order of 1D field storage and its matching to 2D simulation are
important to ensure correct implementation of the 2D TF/SF scheme.
Before updating the 1D field, the code needs to store at each time
instant the 1D field values at the crossing points between line $a$ and
lines $e$, $f$, $g$, and $h$.

The field matching at the TF/SF boundary is performed
differently in accordance with the different updating schemes introduced
in Section \ref{sec2:subsec1}. In the ADE approach, the TF/SF boundary matching
equations on lines $e$, $f$, $b$ read,
\begin{eqnarray}
E^{\rm sca}_x\vert^{n+1}_{i+1/2,j1}&=& E^{\rm sca}_x\vert^{n+1}_{i+1/2,j1}
                      -a_{x2}H^{inc}_{zy}\vert^{n+1/2}_{i+1/2,j1+1/2}, \label{eqn:eqn14}\\
H^{\rm tot}_{zy}\vert^{n+3/2}_{i+1/2,j1+1/2}&=&H^{\rm tot}_{zy}\vert^{n+3/2}_{i+1/2,j1+1/2}
                                   -b_{y2}E^{\rm sca}_x\vert^{n+1}_{i+1/2,j1},\label{eqn:eqn15}\\
H^{\rm tot}_{zx}\vert^{n+3/2}_{i1+1/2,j+1/2}&=&H^{\rm tot}_{zx}\vert^{n+3/2}_{i1+1/2,j+1/2}
                                   -b_{x2}E^{\rm sca}_y\vert^{n+1}_{i1,j+1/2}.\label{eqn:eqn16}
\end{eqnarray}
These updates are performed immediately after Eqs. (\ref{eqn:a}),
(\ref{eqn:c}), and (\ref{eqn:b1}). For the $E_y$-field update on lines
$b^\prime$ and $c^\prime$, because $J_y$ depends on the updated value of $E_y$
in Eq. (\ref{eqn:b3}), the $E_y$ boundary matching is performed
in between Eqs. (\ref{eqn:b2}) and (\ref{eqn:b3}), for example, on
line $b^\prime$,
\begin{equation}
E^{\rm sca}_y\vert^{n+1}_{i1,j+1/2}= E^{\rm sca}_y\vert^{n+1}_{i1,j+1/2}
                      -a_{y2}H^{inc}_{zx}\vert^{n+1/2}_{i1+1/2,j+1/2}. \label{eqn:eqn17}\\
\end{equation}

In the UPML formulation, the TF/SF boundary matching is carried out
immediately after the $P$ and $B$ updates (before updating $D$ and $H$)
in Eqs. (\ref{eqn:l}-\ref{eqn:i}), for example, the updates on lines
$b$ and $b^\prime$ read
\begin{eqnarray}
P^{\rm sca}_y\vert^{n+1}_{i1,j+1/2}&=& P^{\rm sca}_y\vert^{n+1}_{i1,j+1/2}
                      -\alpha_{y1}H^{inc}_{z}\vert^{n+1/2}_{i1+1/2,j+1/2}, \label{eqn:eqn18}\\
B^{\rm tot}_{z}\vert^{n+3/2}_{i1+1/2,j+1/2}&=&B^{\rm tot}_{z}\vert^{n+3/2}_{i1+1/2,j+1/2}
                                   -\beta_{2}E^{\rm sca}_y\vert^{n+1}_{i1,j+1/2}.\label{eqn:eqn19}
\end{eqnarray}
Because the above updates are performed between the updates of
$P$ and $D$ or $B$ and $H$, they are indicated in the flow chart (Fig. \ref{fig:f2})
by the upward arrows on the right. We note that if the same type of
PML absorbing boundary condition is used to terminate both the 1D and the 2D
field propagation, one can allow them to have the same updating coefficients
in the PML region and therefore remove the procedures
of saving and matching the field components on lines
$g$ and $h$ [$E_{\rm BOT}$ and $H_{\rm BOT}$].\cite{Jiang}
This particular setup is useful in the description of a very
thick bottom layer (semi-infinite in
the positive $y$ direction).

In the case of normal incidence,
the code simplifies in two ways. First,
in Eq. (\ref{eqn:eqn7}), $H=H^\prime$,
and therefore Eq. (\ref{eqn:eqn11}) is removed from the 1D $H$-field
update procedure. Second, it is not necessary to store and interpolate the
field values $E_{\rm BOT}$, $E_{\rm TOP}$, $H_{\rm BOT}$, and $H_{\rm TOP}$,
because field excitation is synchronized
along lines $e$, $f$, $g$ and $h$, respectively.

The incident field values on lines $a$, $b$, $c$ and $d$ are calculated
from Eqs. (\ref{eqn:eqn8}-\ref{eqn:eqn11}) using a 1D TF/SF scheme that allows
fields reflected from the interfaces to exit the 1D simulation domain.
Based on the geometry shown in the lower left panel in Fig. \ref{fig:f1},
we assume that the incoming $H$-field with time-dependence
$f(t)$ excites the 1D field at point $(i1-1/2,j1-3/2)$. Paired
with this excitation is an $E$-field of the
form $g(t)=-\sqrt{\mu_0/\epsilon_0}f(t+\Delta x\cos(\theta)/2c)\cos(\theta)$,
exciting the 1D field at point $(i1-1/2,j1-2)$. For example,
on line $a$ in Fig. \ref{fig:f1}, the 1D TF/SF boundary matching equations read,
\begin{eqnarray}
E^{\rm sca}_a\vert^{n+1}_{j1-2}&=& E^{\rm sca}_a\vert^{n+1}_{j1-2}
                      -a_{x2}f\vert^{n+1/2}, \label{eqn:eqn20}\\
H^{\rm \prime tot}_a\vert^{n+3/2}_{j1-3/2}&=&H^{\rm \prime tot}_a\vert^{n+3/2}_{j1-3/2}
                                   -b_{y2}g\vert^{n+1}.\label{eqn:eqn21}
\end{eqnarray}
The values of $f\vert^{n+1/2}$ and $g\vert^{n}$ are
calculated from the known expressions of $f(t)$ and $g(t)$ using
time-domain interpolation when necessary. These values are stored
at each instant to generate the excitation fields for lines
$b$, $c$, and $d$ by introducing a time delay $t_{delay}=N\Delta x\sin(\theta)/c$.
The field values on the horizontal lines $e$, $f$, $g$ and $h$ are obtained in a
similar fashion.
One can also store the field values at each point on line $a$, and save the
computation along lines $b$, $c$ and $d$ by introducing a proper time delay.
This scheme reduces the computation time for the cost of larger memory requirement.
Finally, the 2D TF/SF boundary values [$E^{\rm inc}_y$]
along lines $b^\prime$ ($c^\prime$) are readily calculated from the $H$-field
values on lines $a$ and $b$ ($c$ and $d$) using
Eqs. (\ref{eqn:b2}) and (\ref{eqn:b3}). In addition, we note that the
excitation and PML absorbing boundary conditions
are enforced on $H^\prime$ in the 1D field updates.

Several practical issues should be considered. First,
in vacuum, the projection of the phase velocity of the oblique
incident field on the $y$-axis is $c/\cos(\theta)$. As the incident angle
$\theta$ increases, the phase velocity can be very large and
cause numerical instability if a fixed Courant criterium is enforced (e.g.,
$\Delta t=\Delta x/2c$). Based on this observation, we vary the
Courant number $S=c\Delta t/\Delta x$ to ensure stability. When
the incident angle is small, a small Courant number is  used
to ensure resolution of the time domain interpolation along the
horizontal boundaries. In our simulation, the same Courant number is used
for both 1D and 2D wave propagations, while an interpolation scheme
to match different Courant numbers in 1D and 2D wave propagations is
explained in Ref. \onlinecite{Capoglu}. Second, as the dielectric
function is discontinuous across the interface between
layers of different media, we have used an average dielectric function for
updating the fields at the interface.\cite{Zhao,Mohammadi1} For example,
in the left panel of Fig. \ref{fig:f1}, we use the dielectric function
$\epsilon_{\rm eff}=(\epsilon_{1r}+\epsilon_{2r})/2$ for
the $E_x$-field updates at the interface.
In Section \ref{sec:sec3}, we will show that this scheme leads
to faster convergence and/or higher accuracy as compared to the standard step-like change of $\epsilon$. Finally, we use
a Gaussian ramping in the hard source time-response in
Eqs. (\ref{eqn:eqn20}) and (\ref{eqn:eqn21}) to slowly ramp the field
to continuous wave so as to avoid high-frequency component leakage
out of the TF/SF domain. Specifically,
$f(t) = \exp(-(t-\tau_{dealy})^2/\tau_0^2)\sin(\omega t)$,
$\tau_{delay}=30$ fs, $\tau_0=10$ fs, for the ramping
phase $0<t<\tau_{delay}$.

\subsection{Numerical tests}

To test the accuracy of the TF/SF scheme, we compare our simulation results to
analytical results by considering the oblique TM wave incident
upon a slab sandwiched between two vacuum media. The analytical results
for the reflection and transmission coefficients are given by\cite{Wolf}
\begin{eqnarray}
r&=&\frac{r_{12}+r_{23}e^{2i\gamma}}{1+r_{12}r_{23}e^{2i\gamma}},\label{eqn:eqn22}\\
t&=&\frac{t_{12}t_{23}e^{i\gamma}}{1+r_{12}r_{23}e^{2i\gamma}}.\label{eqn:eqn23}
\end{eqnarray}
For TM wave,
\begin{eqnarray}
r_{ij}&=&\frac{\cos(\theta_i)/n_i -\cos(\theta_j)/n_j}{\cos(\theta_i)/n_i+\cos(\theta_j)/n_j},\label{eqn:eqn24}\\
t_{ij}&=&\frac{2\cos(\theta_i)/n_i}{\cos(\theta_i)/n_i+\cos(\theta_j)/n_j},\label{eqn:eqn25}\\
\sin(\theta_i)&=&\frac{n_j}{n_i}\sin(\theta_j),\label{eqn:eqn26}\\
\gamma&=&\frac{\omega}{c}n_2h\cos(\theta_2).\label{eqn:eqn27}
\end{eqnarray}
Here, $r_{ij}$ and $t_{ij}$ denote, respectively, the reflection and transmission
coefficients at the interface between media $i$ and $j$, and
$n_i=\sqrt{\epsilon_i(\omega)}$ denotes the refractive index of media $i$.
We assume that medium $2$ is a slab of thickness $h$.
The waves at the input side of the slab, where the incident and reflected waves propagate, and at the
output side, where the transmitted wave propagates, can then be expressed as,
\begin{eqnarray}
\psi^{\rm input}&=&\exp[i(k_x x+k_y y)]+r\exp[i(k_x x-k_y y)],\label{eqn:eqn28}\\
\psi^{\rm output}&=&t\exp[i(k_x x+k_y y)].\label{eqn:eqn29}
\end{eqnarray}
The above expressions indicate that the maximum field amplitude on
the input and output sides are $1+|r|$ and $|t|$, respectively. These quantities
can be obtained along a $y$-direction detection line in the TF/SF scheme
for layered media without placing any scatterer inside the TF region. Using this
scheme, we also test the leakage, defined as the ratio of the maximum field magnitude in
the scattered field region to the maximum field magnitude in the total
field region:
${\rm leakage} = {\rm max}|\psi^{\rm sca}\vert^{n}_{i,j}|/{\rm max}|\psi^{\rm tot}\vert^{n}_{i,j}|$,
where $\psi$ refers to $E_x$, or $E_y$, or $H_z$.
In the ideal case, ${\rm leakage} = 0$, whereas in practice, ${\rm leakage} <10^{-2}$
(or $-40$ dB) is desirable.\cite{Taflove} Specific
numerical examples of the tests are provided in Section \ref{sec:sec3},
where ``leakage'' refers to the largest leakage among $E_x$, or $E_y$, or $H_z$.
In addition, we
have tested the accuracy of the wave propagation in the layered media by inspecting the
$x$ and $y$ projections of the wavelength
[where, for instance, the $x$-projection
is $2\pi/k_x$ and $k_x=k\sin(\theta)$]. In the case of a dispersive slab,
we have also tested the skin depth
(the distance
where the field decays to $e^{-1}$
of its value at the surface, ca. 30 nm for the Drude model
and parameters in our calculation), by considering a slab with thickness
larger than $300$ nm. These tests all show an error within $5\%$
compared to analytical results.

\section{Numerical examples and discussions}
\label{sec:sec3}

\begin{figure}[htp]
\begin{center}
\includegraphics[width=12.5cm]{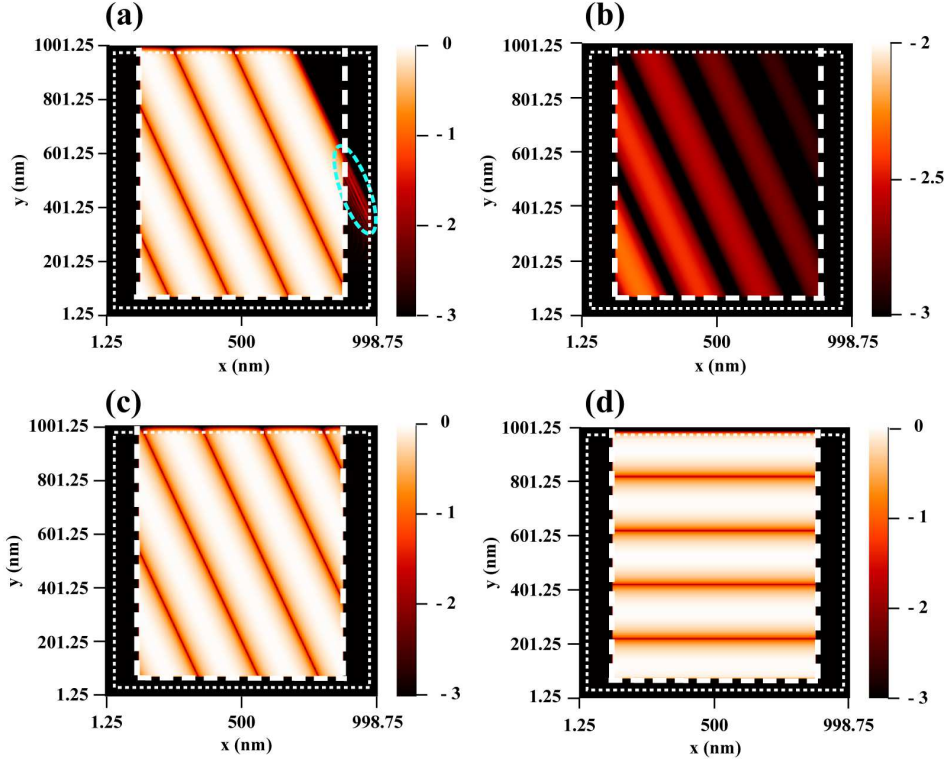}
\caption{(color online) In panels (a--c), the incident plane wave
enters the TF region from the lower left corner of
the TF/SF boundary with incident angle
$\theta=65^\circ$. (a) Magnetic ($H$) field snapshot
at $3.00$ fs for a plane wave propagating in vacuum.
The dashed oval indicates the leakage outside
the TF/SF boundary as a
result of the instantaneous turn-on of the field.
Panels (b--d) show $H$-field snapshots for
a plane wave propagation with initial Gaussian
ramping. (b) $H$-field snapshot during ramping
(at $7.34$ fs) and (c) after steady state is established
(at $66.71$ fs). (d) $H$-field snapshot at $90.06$ fs
for a plane wave propagating in the positive $y$ direction.
For all calculations,
the incident wavelength is $\lambda=400$ nm
(period $T=1.33$ fs), and the steady-state amplitude
of the incident magnetic field is $1$ A/m. The mesh
size is $\Delta x=2.5$ nm, and the Courant number is $S=0.4$.
A log color scale ($\log_{10}|H_z|$) is used in all plots. The thick, dashed
(thin, dotted) rectangle indicates the TF/SF (inner PML) boundary.}
\label{fig:fig1}
\end{center}
\end{figure}

To illustrate the generality of our formulation, we first consider
the simple case of plane wave propagation in vacuum, illustrated
in Fig. \ref{fig:fig1}. In panels (a-c), the plane wave
(wavelength $400$ nm) is injected from the lower left corner
into the TF region (bounded by the thick, dashed lines) with incident angle
$\theta=65^\circ$. In panel (d), the plane wave propagates in
the positive $y$ direction. It is shown that as the field penetrates into
the Berenger PML located at the top of the simulation domain,
it is efficiently absorbed. Negligible leakage is introduced at the
PML boundary as the 1D field updating equations acquire the
same coefficients as the 2D equations [see Section \ref{sec:sec2}].
The dashed oval in panel
(a) indicates considerable leakage ($3.292\times10^{-2}$)
outside the TF region because in this case the incident continuous
wave (cw) field is turned on instantaneously. Consequently,
the high frequency components in the leading wave front are not well
matched at the TF/SF boundary, resulting in the leakage.
As shown by panels (b) and (c), the leakage can be reduced by one
order of magnitude by slow (Gaussian) ramping of the incident
field to steady-state cw oscillations. In light of this,
hereafter we use Gaussian ramping prior to cw in the excitation
hard source $f(t)$ and $g(t)$. The maximum leakage  in the calculations
of panels (b--d) is $1.367\times10^{-3}$ and the relative
error in the vacuum wave impedance
($Z_0=\sqrt{\mu_0 / \epsilon_0}$) is $0.41\%$.

\begin{figure}[htp]
\begin{center}
\includegraphics[width=12.5cm]{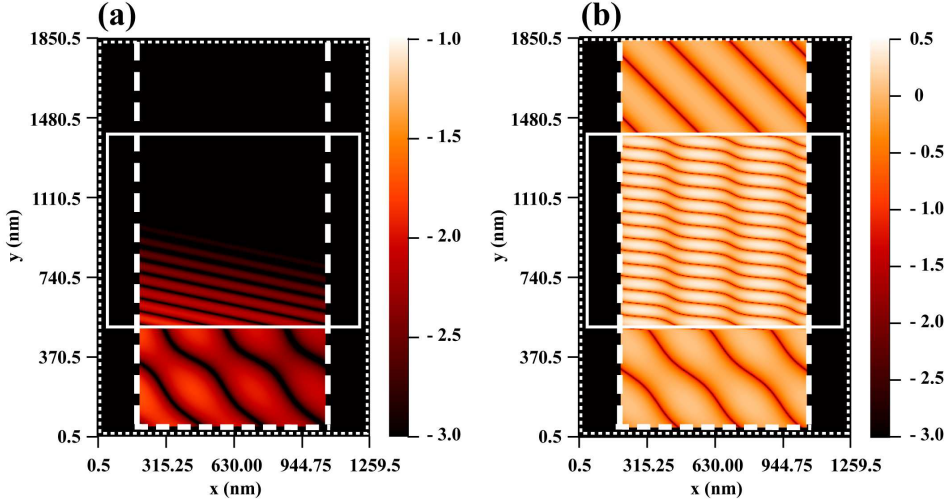}
\caption{(color online) Magnetic field snapshots of a
plane wave obliquely incident on a dielectric slab
(indicated by a solid rectangle).
Reflection and refraction (a) at the lower interface
(at $10.01$ fs) and (b) after steady state is established
(at $100.07$ fs). The incident plane wave enters the
TF region from the lower left corner of the TF/SF
boundary with incident angle
$\theta=45^\circ$. The
incident wavelength in vacuum is $\lambda=400$ nm
(period $T=1.33$ fs), and the steady-state amplitude
of the incident magnetic field is $1$ A/m in all calculations. The
dielectric constant and thickness of the slab
are $\epsilon_r=11.7$ and $900$ nm, respectively.
The media above and below the slab are vacuum, the
mesh size is $\Delta x=1.0$ nm, and the Courant number is $S=0.3$.
A log color scale is used in all plots. The thick, dashed
(thin, dotted) rectangle indicates the TF/SF (inner PML) boundary.
The slab does not penetrate into the PML region.}
\label{fig:fig2}
\end{center}
\end{figure}

As a second example, we study a plane wave obliquely incident on a
dielectric slab.\cite{Winton, Capoglu}. In Fig. \ref{fig:fig2}, we
plot  snapshots of the magnetic field of a plane wave
(wavelength $400$ nm)  incident at an angle $\theta=45^\circ$
on a $900$-nm thick dielectric slab (dielectric constant
$\epsilon_r=11.7$). In both panels, the solid rectangle
indicates the location of the slab,
while the thick, dashed rectangle shows the TF/SF boundary.
The plane wave is injected from the lower left corner and first
impinges on the lower vacuum/dielectric interface.
In panel (a), we observe the interference patterns of the
reflected wave with the incident wave below the lower
vacuum/dielectric interface while the refracted wave front
 propagates in the slab. The faint wave front in the
dielectric slab is due to the slow Gaussian ramping of the
incident field. After a steady state is established
[Fig. \ref{fig:fig2}(b)], the magnetic field pattern
clearly reveals the interference between the reflected
and incident waves, the interference within the dielectric slab,
and the final transmission through the slab.
In Fig. \ref{fig:fig2} (b), it is observed that the final
transmitted wave maintains the same propagation
direction  as the incident
wave ($45^\circ$ to $+y$ direction)
because the media below and above the slab are both vacuum.

\begin{figure}[htp]
\begin{center}
\includegraphics[width=12cm]{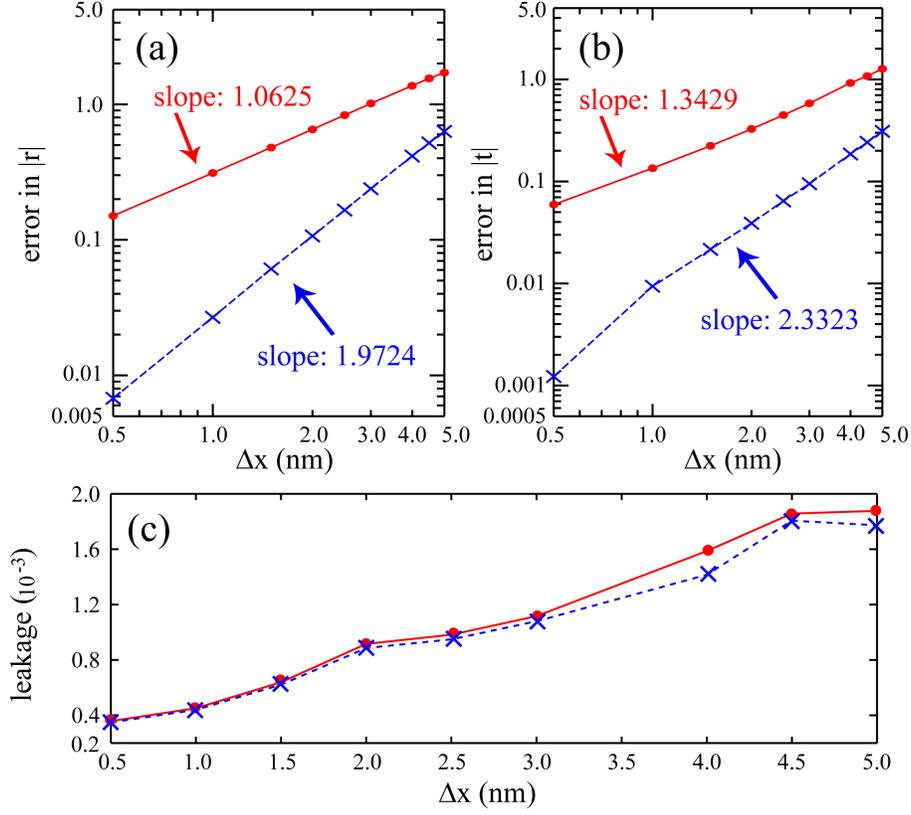}
\caption{(color online)
Relative error in the magnitude of (a) the
reflection coefficient $r$ and (b) the transmission
coefficient $t$ as a function of the mesh size $\Delta x$.
(c) Maximum leakage as a function of mesh
size $\Delta x$. In all calculations, the Courant
number is $S=0.3$. In all figures, the red, solid
(blue, dashed) curve shows the result without (with)
the interface averaging of the dielectric constants.
The parameters of the incident wave and the dielectric
slab are as in the calculation leading to Fig. \ref{fig:fig2}.}
\label{fig:fig3}
\end{center}
\end{figure}

We proceed to examine the convergence of the magnitude of the reflection ($r$) and transmission ($t$) coefficients to  the analytical results given by Eqs. (\ref{eqn:eqn22}) and (\ref{eqn:eqn23}).
In Fig. \ref{fig:fig3}, we plot the relative errors in (a)
$|r|$ and (b) $|t|$ with respect to analytical results as a
function of the mesh size $\Delta x$. The red, solid (blue, dashed)
curve in Fig. \ref{fig:fig3} shows the convergence result
without (with) the interface averaging of the dielectric
constant. From the comparison, it is clear that
calculations with interface correction lead to uniformly
smaller error than that without the interface correction.
The slope of each line in the log-log plot obtained by
the least-square fit indicates that second order
accuracy of Yee's algorithm is maintained with the
interface correction, while the accuracy degrades to
first order without the interface correction.
Similar effects have been reported in previous studies
on the accuracy of FDTD results with dielectric
interfaces,\cite{Hirono,Hwang} while here we observe such effects
within the TF/SF formulation in the context of
layered media. We note that the interface correction
scheme does not entail additional computational and
memory requirements and is thus always recommended.
In Fig. \ref{fig:fig3} (c),
we show that the maximum leakage with interface averaging
is uniformly smaller than that without the interface
averaging for different mesh sizes. Throughout, the
maximum leakage is below $2.0\times10^{-3}$,
substantiating our confidence in the
TF/SF scheme.\cite{Taflove}

To further test the accuracy of the dielectric slab
reflection and transmission upon oblique incidence
of a plane wave, we compare the analytical results
with FDTD calculated results at different incident
wavelengths in Table \ref{tab:tab1} and at different
incident angles in Table \ref{tab:tab2}. As shown,
the relative error (given in parentheses) is uniformly
below $5\%$, except for incident angle
$\theta=50^{\circ}$, where $|r|$ is below $0.01$.
It is interesting to note that the relative error
diminishes with increasing wavelength, while a
non-monotonic trend is seen in the errors of both
reflection and transmission coefficients for an
increasing incident angle.

\begin{figure}[htp]
\begin{center}
\includegraphics[width=12.5cm]{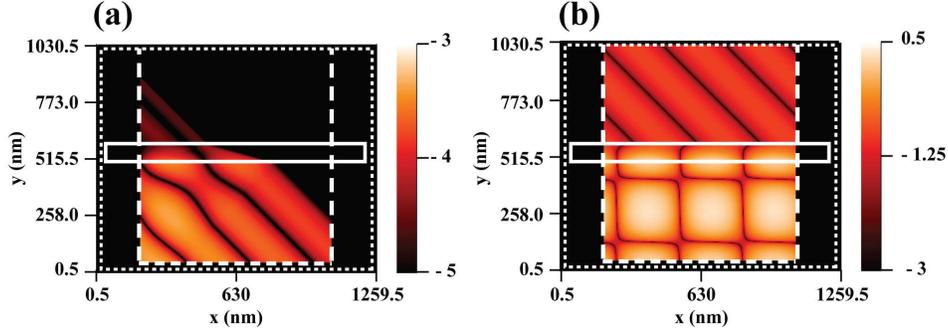}
\caption{(color online) Magnetic field snapshots of a
plane wave obliquely incident on a Drude metal slab
(indicated  by a solid rectangle).
Reflection, refraction and transmission
(a) before
(at $2.50$ fs) and (b) after  (at $60.04$ fs) steady state is
established. In all calculations, the
incident plane wave enters the TF region from the
lower left corner of the TF/SF boundary (dashed lines)
with incident angle
$\theta=45^\circ$. The incident wavelength
in vacuum is $\lambda=400$ nm (period $T=1.33$ fs),
and the steady-state amplitude of the incident magnetic
field is $1$ A/m. The metal slab is $80$ nm thick with
Drude parameters: $\epsilon(\infty)=7.0246$,
$\omega_D=1.5713\times10^{16}$ rad/s,
and $\Gamma_D=1.4003\times10^{14}$ rad/s. The media
above and below the slab are vacuum, the mesh size is $\Delta x=1.0$ nm,
and the Courant number is $S=0.3$.
A log color scale is used in all plots. The thick, dashed
(thin, dotted) rectangle indicates the TF/SF (inner PML) boundary.
The slab does not penetrate into the PML region.}
\label{fig:fig4}
\end{center}
\end{figure}

Next we apply the TF/SF method to study the
reflection and refraction of a plane wave
obliquely incident upon a dispersive metal slab.
Snapshots of the magnetic field are shown
in Fig. \ref{fig:fig4} as the plane wave passes
through the metal slab. Specifically, we consider
an incident plane wave with wavelength $400$ nm
and $\theta=45^\circ$ injected from the lower
left corner of the TF region upon an $80$ nm thick
dispersive metal slab described by the Drude
model $\epsilon_m=\epsilon(\infty)-\omega_D^2/(\omega^2+i\Gamma_D\omega)$,
with $\epsilon(\infty)=7.0246$,
$\omega_D=1.5713\times10^{16}$ rad/s, and
$\Gamma_D=1.4003\times10^{14}$ rad/s.
This set of parameters is optimized to fit the dielectric
data reported in Ref. \onlinecite{Johnson} for bulk silver in
the spectral range from $330$ to $500$ nm.
Figures \ref{fig:fig4} (a) and (b) illustrate the
magnetic field distribution before and after reaching
a steady state, respectively. In
Fig. \ref{fig:fig4} (b), the large curvatures at the
interference minima between the incident and reflected
fields below the lower interface indicate a large
reflection coefficient ($>0.9$). Inside the metal, because
of the complex dielectric function of the slab, the wave
front is no longer a plane wave, as is clearly discernable
in Fig. \ref{fig:fig4}. However, the final transmitted
wave exiting from the upper interface recovers a plane
wave front and the same propagation constant as the
incident wave, because the media below and above the
dispersive slab are both vacuum. From
Figs. \ref{fig:fig2}(b) and \ref{fig:fig4}(b),
it is seen that the periodicity
in the $x$ direction of the fields below, inside,
and above the slab is the same. By further observing
the field propagation after reaching the steady state
in both cases (not shown), it is clear that the phase of the total
field in the $x$ direction is matched. This observation
confirms the phase matching condition parallel to the
interface (same $k_x$ across the interfaces), which
is critical to the derivation of the 1D field propagation,
Eq. (\ref{eqn:eqn5}).

\begin{figure}[htp]
\begin{center}
\includegraphics[width=12cm]{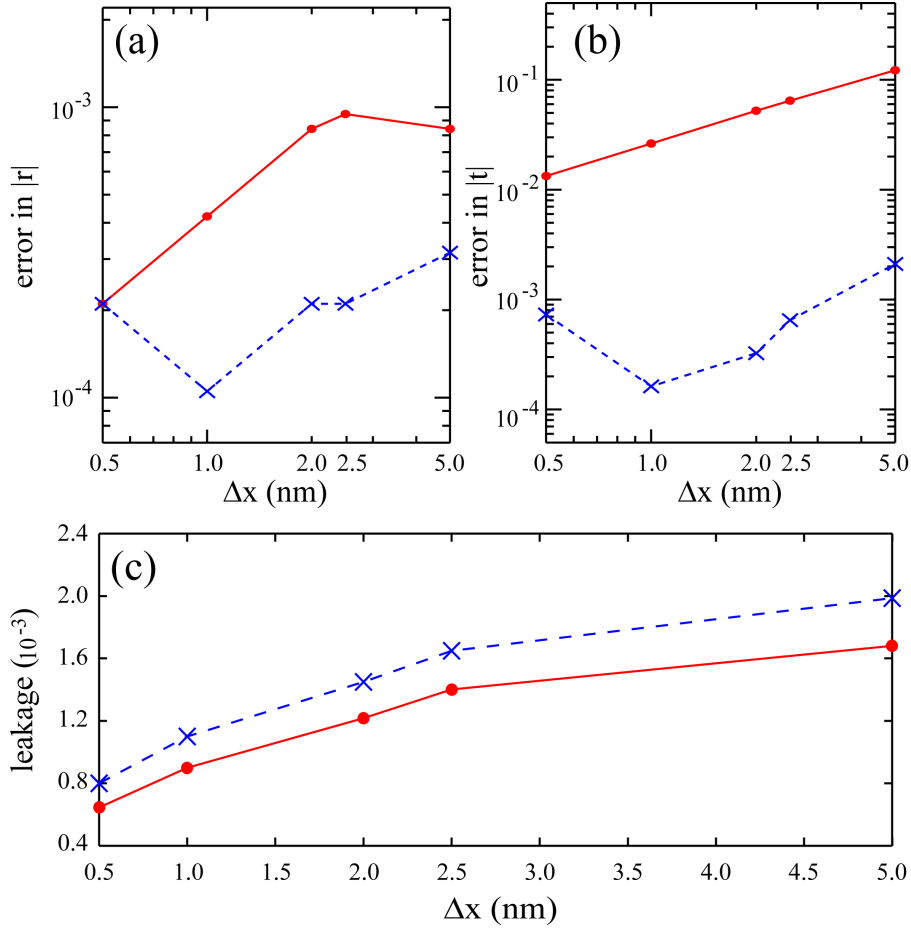}
\caption{(color online) Relative error in the magnitude
of the (a) the reflection coefficient $r$ and (b) the transmission
coefficient $t$ as a function of the mesh size $\Delta x$.
(c) Maximum leakage as a function of mesh
size $\Delta x$. The Courant number is $S=0.3$  in all calculations.
In all figures, the red, solid (blue, dashed) curve
shows the result without (with) the interface averaging
of the dielectric constants. The parameters of the incident
wave and the metal slab are the same as in the calculations leading to
 of Fig. \ref{fig:fig4}.}
\label{fig:fig5}
\end{center}
\end{figure}

To examine the convergence of our results in
the case of the metal slab, we use the same
incident field condition as that in Fig. \ref{fig:fig4}
and plot the
relative error of the reflection
and transmission coefficients
as a function of the mesh size
$\Delta x$ in Figs. \ref{fig:fig5} (a) and
(b), respectively. It is seen that the results
with interface averaging (blue, dashed curves) of the dielectric
function yield uniformly
lower error than the results without the
interface averaging (red, solid curves).
A first-order
power law is seen in the error of the
transmission coefficient as a function of
the mesh size without interface averaging,
all other errors are near and below $10^{-3}$,
illustrating the convergence of the FDTD results.
FDTD simulations on similar dispersive systems
have been reported by Mohammodi et al.,
who suggested that the dispersive contour-path method is
able to achieve smaller error even for a relatively
large step-size ($\Delta x$).\cite{Mohammadi2}
Fig. \ref{fig:fig5}(c) shows that the leakage
decreases with a decreasing mesh size,
albeit in this case the leakage with the interface
averaging of the dielectric function is slightly
larger than that without the
averaging [cf. Fig. \ref{fig:fig3}(c)].

In Tables \ref{tab:tab3} and \ref{tab:tab4}, we
compare between the analytical and
the FDTD calculated reflection and transmission coefficient magnitudes at
various incident wavelengths and incident angles
 for the
metal slab studied in Fig. \ref{fig:fig4}. The
FDTD results are obtained after steady state
is reached under cw incident plane wave illumination.
In the frequency domain, this corresponds to a
fixed incident wavelength, and the
Drude model provides a constant complex value
of dielectric function, which can be used in
Eqs. (\ref{eqn:eqn22}) and (\ref{eqn:eqn23}) to obtain
the reflection and transmission coefficients.
In Table \ref{tab:tab3}, we  list the free space wavelength in the
 $350$ to $500$ nm range, to which the fitted
Drude model is applicable. The small relative
errors ($<2.5\%$) shown in the parentheses in
Tables \ref{tab:tab3} and \ref{tab:tab4} illustrates the reliability of
our calculations using the TF/SF formulation in the
case of  layered dispersive media. The maximum
leakage found in obtaining the data in Tables \ref{tab:tab3}
and \ref{tab:tab4} is $1.201\times10^{-2}$,
which occurs at $\theta=80^{\circ}$.

\begin{figure}[htp]
\begin{center}
\includegraphics[width=8.5cm]{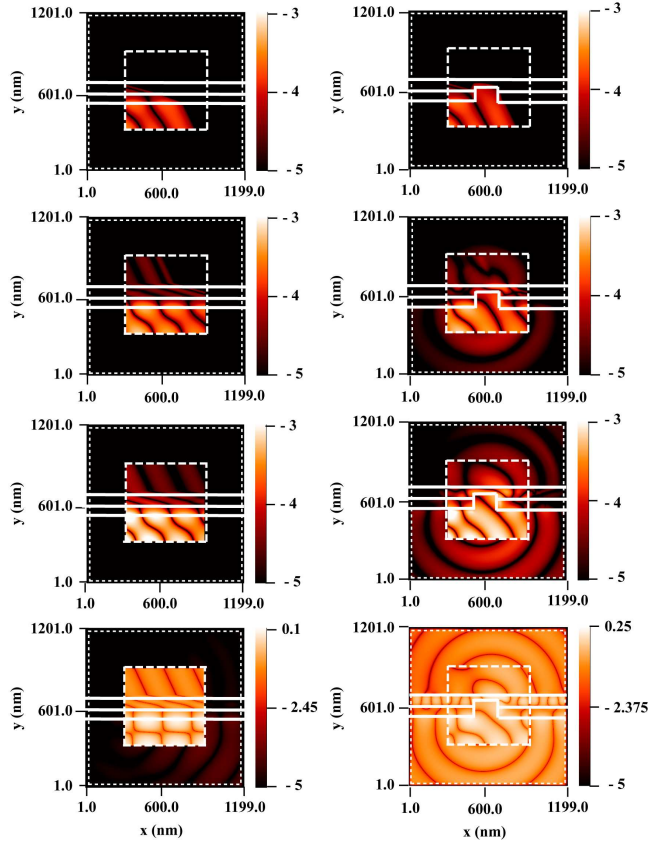}
\caption{(color online) Magnetic field snapshots for a
plane wave obliquely incident on two layers of materials.
The interfaces between the layers and vacuum are indicated
by solid horizontal lines. The lower layer is an $80$ nm
thick Drude metal with parameters: $\epsilon(\infty)=7.0246$,
$\omega_D=1.5713\times10^{16}$ rad/s, and
$\Gamma_D=1.4003\times10^{14}$ rad/s. The upper layer is
a $100$ nm dielectric with dielectric constant
$\epsilon_r=11.7$. The media below and above the two
layers are vacuum. The panels in the right column
illustrate the scattering due to a slit of width $200$ nm and
depth $120$ nm in the same layered
structure as in the left column. Rows 1, 2, and 3 show
snapshots at $1.60$, $3.20$, $4.80$ fs (before a steady
state is established); Row 4 shows the snapshot at $60.04$ fs
(after a steady state is established). For all calculations,
the incident plane wave enters the TF region from the lower
left corner of the TF/SF boundary with with incident angle
$\theta=65^\circ$. The incidence wavelength is
$\lambda=400$ nm (period $T=1.33$ fs), and the steady-state
amplitude of the incident magnetic field is $1$ A/m. The mesh
size is $\Delta x=2.0$ nm, and the Courant number is $S=0.3$.
A log color scale is used in all plots. The thick, dashed
(thin, dotted) rectangle indicates the TF/SF (inner PML) boundary.
The slabs are extended into the UPML region.}
\label{fig:fig6}
\end{center}
\end{figure}

Panels in the left column of Fig. \ref{fig:fig6}
illustrate snapshots of the magnetic field
as the wave propagates through two-layer media
consisting of a lower layer of $80$-nm thick dispersive
material and an upper layer of $100$-nm thick dielectric
material under oblique plane wave incidence.
The material parameters are given in the caption
of Fig. \ref{fig:fig6}. In these panels, the solid
horizontal lines define the boundaries between
different layers, which are extended into the UPML
in the $x$ direction. The dashed box denotes
the TF/SF boundary.  The incident plane wave with
wavelength $400$ nm and $\theta=65^\circ$ is injected
from the lower left corner of the TF region.
The magnetic field snapshots in the first column of
Fig. \ref{fig:fig6} show that Snell's law is obeyed
when the field passes through the two layers of
materials. In particular, the propagation direction
in the high-index dielectric material
exhibits a smaller angle to the normal than
the incident wave, whereas the final transmitted
wave  propagates along the direction of the
incident wave. More importantly, we observe that
the phase of the waves across the different layers
is matched in the $x$ direction after a steady
state is established (bottom panel in the left column),
which is again consistent with Eq. (\ref{eqn:eqn5}).
In this case, the magnitude of the reflection and
transmission coefficients calculated by FDTD is
$|r|=0.9524$ and $|t|=0.0866$, respectively.
The bottom panel in the left column also shows
non-negligible leakage penetrating through the TF/SF
box and propagating into the lower right corner
of the simulation domain, nevertheless, the maximum value of
the leakage in $H_z$
is $1.5461\times10^{-4}$, which is
insignificant in practice.

Panels in the second column of Fig. \ref{fig:fig6}
are obtained under the same conditions as those in
the first column except that a slit
of $200$ nm width (in the $x$ direction)
and $120$ nm depth (in the $y$ direction) is placed
in the middle of the simulation domain. In the
TF region, the slit causes strong scattering of
the injected plane wave, which results in the
observed interference patterns. The slit introduces entirely new physics:
outside the TF/SF boundary, the purely scattered
wave distribution is
reminiscent of a dipole radiation pattern. Closer
inspection reveals that the field distribution is
asymmetric with respect to the slit center ($x=600$ nm).
The scattered field is strongest near the lower
surface of the dispersive slab and to the right of
the TF/SF box and weakest above the upper surface
of the dielectric slab and to the left of the TF/SF box.
The asymmetric angular distribution is a clear signature
of the oblique incidence of the exciting
plane wave. By enlarging
the SF region size, we find that
the purely scattered wave along the
lower surface of the metal thin film
consists mainly of
surface plasmon polariton (SPP) waves propagating
away from the slit. These are identified by
their wavelength - $~348$ nm in the $x$ direction compared with
the analytical result for the wavelength of SPP at
the interface between vacuum and metal, which
is given by
$\lambda_0/\rm{Re}\sqrt{\epsilon_m/(1+\epsilon_m)} =348.25$ nm.

For the simulations in Fig. \ref{fig:fig6}, we have
updated the field at the horizontal and vertical
interfaces using the averaging
scheme discussed above, and have tested the convergence of the fields in the TF and SF regions
with respect to mesh size ($\Delta x$), TF box size, and
physical size of the simulation region. We note that
UPML termination of the simulation domain is
important because the scattered field due to
the slit is significant. Our tests show that the maximum
scattered field in the SF region is only one order
of magnitude less than the maximum field in the TF region.
Additionally, the UPML can effectively absorb the outgoing wave
in the dispersive and dielectric layers.
Furthermore, the boxed TF/SF boundary has
advantage over the
$\Pi$-shaped boundary considered previously,
particularly when one is interested in the full
angular distribution of the scattered field
in the far-field zone.




\section{Conclusions}
\label{sec:sec4}

Using Maxwell's equations for the transverse magnetic wave, along with translational invariance
and phase matching principles,
we derived an equivalent
one-dimensional wave propagation equation along the direction perpendicular
to the interfaces between layered media. We then derived the corresponding
finite-difference time-domain equations for
layered dielectric media and dispersive media with a Drude pole pair.
To utilize these equations for a plane wave with oblique incidence,
we discussed the simulation setup and procedure in the framework of the total
field / scattered field formulation with a special emphasis on techniques
to match the fields at the total field / scattered field boundary. We
have performed tests on vacuum propagation and on
the reflection and refraction at a dielectric and a dispersive slab.
Converged simulation
results for various incident angles and wavelengths reveal that the errors in the reflection and refraction
coefficients are uniformly below $5\%$ compared to analytic results.
The numerical example of scattering at
a nano-scale (sub-wavelength) slit in a dispersive medium invites interesting applications of our formulation to
time-dependent studies of electromagnetic wave scattering
at surface or embedded scatters in dispersive media, for example,
the coupling of incident oblique plane wave into surface plasmon
polaritons. For this purpose, the developed method offers the flexibility
of choosing the total field region inside which the near-field exhibits
interference pattern between incident and scattered fields, while outside
which the scattered far-field can be detected at all angles.

\begin{acknowledgments}
This work is supported by the W. M. Keck Foundation (grant number 0008269) and by the National Science Foundation (grant number ESI-0426328). The authors thank
Hrvoje Petek and Atsushi Kubo for the communications of corresponding
experimental data, and Maxim Sukharev, Gilbert Chang,
Jeffrey McMahon, Stephen Gray and Allen Taflove for insightful
discussions. They are particularly thankful to \.{I}lker \c{C}apo\v{g}lu for
discussions on the total field/scattered field formalism.
\end{acknowledgments}

\appendix
\section{TE mode wave propagation}
\label{Appx1}

Maxwell's equations in 2D in the
frequency domain for the TE mode read,
\begin{gather}
\frac{\partial H_{y}}{\partial x}-\frac{\partial H_{x}}{\partial y}
=-i\omega\epsilon_{0}\epsilon(\omega)E_{z},\label{eqn:eqn30}\\
\frac{\partial E_{z}}{\partial y}=i\omega\mu_0H_{x},\label{eqn:eqn31}\\
\frac{\partial E_{z}}{\partial x}=-i\omega\mu_0H_{y}.\label{eqn:eqn32}
\end{gather}
Substituting Eq. (\ref{eqn:eqn32}) into Eq. (\ref{eqn:eqn30}) yields,
\begin{equation}
\frac{\partial H_{x}}{\partial y}
=i\omega\epsilon_{0}\left[\epsilon(\omega)-\epsilon_{1r}\sin^2(\theta)\right]E_{z},\label{eqn:eqn33}\\
\end{equation}
where $\epsilon_{1r}$ denotes the relative permitivity of the first
medium (see Fig. \ref{fig:f1}). Equations (\ref{eqn:eqn31}) and (\ref{eqn:eqn32})
are used for 1D TE mode wave propagation. These equations can be readily
solved using the same FDTD procedure as for Eqs. (\ref{eqn:eqn1}) and (\ref{eqn:eqn2}).
The time-domain solution is facilitated by the fact that material dispersiveness introduces
a factor $\left[\epsilon(\omega)-\epsilon_{1r}\sin^2(\theta)\right]$ in
Eq. (\ref{eqn:eqn33}),
whereas  in Eq. (\ref{eqn:eqn5}) it introduces a factor
$\left[\epsilon(\omega)-\epsilon_{1r}\sin^2(\theta)\right]/\epsilon(\omega)$, which entails more difficulty for the FDTD solution.\cite{footnote3} The simulation setup
and flow chart in Section \ref{sec:sec2} can be used for the TE mode by
exchanging the roles of the $E$ and $H$ fields.

\section{FDTD equations for 2D TM mode wave propagation}
\label{Appx2}

The FDTD equations based on the auxiliary differential equation (ADE) approach read,
\begin{eqnarray}
E_x\vert^{n+1}_{i+1/2,j}&=& a_{x1}E_x\vert^{n}_{i+1/2,j}
                      +a_{x2}(H_{zy}\vert^{n+1/2}_{i+1/2,j+1/2}-H_{zy}\vert^{n+1/2}_{i+1/2,j-1/2})
                      +a_{x3}J_x\vert^{n+1/2}_{i+1/2,j}, \label{eqn:a}\\
J_x\vert^{n+3/2}_{i+1/2,j}&=&a_{x4}J_x\vert^{n+1/2}_{i+1/2,j}+a_{x5}E_x\vert^{n+1}_{i+1/2,j},\label{eqn:b}\\
E_y\vert^{n+1}_{i,j+1/2}&=& a_{y1}E_y\vert^{n}_{i,j+1/2}
                      +a_{y2}(H_{zx}\vert^{n+1/2}_{i+1/2,j+1/2}-H_{zx}\vert^{n+1/2}_{i-1/2,j+1/2})
                      +a_{y3}J_y\vert^{n+1/2}_{i,j+1/2}, \label{eqn:b2}\\
J_y\vert^{n+3/2}_{i,j+1/2}&=&a_{y4}J_y\vert^{n+1/2}_{i,j+1/2}+a_{y5}E_y\vert^{n+1}_{i,j+1/2},\label{eqn:b3}\\
H_{zx}\vert^{n+3/2}_{i+1/2,j+1/2}&=&b_{x1}H_{zx}\vert^{n+1/2}_{i+1/2,j+1/2}
                                   +b_{x2}(E_y\vert^{n+1}_{i+1,j+1/2}-E_y\vert^{n+1}_{i,j+1/2}),\label{eqn:b1}\\
H_{zy}\vert^{n+3/2}_{i+1/2,j+1/2}&=&b_{y1}H_{zy}\vert^{n+1/2}_{i+1/2,j+1/2}
                                   +b_{y2}(E_x\vert^{n+1}_{i+1/2,j+1}-E_x\vert^{n+1}_{i+1/2,j}),\label{eqn:c}\\
H_z&=&H_{zx}+H_{zy}.
\end{eqnarray}
The coefficients in the $E$-field updating equations
are medium dependent; specifically,
in vacuum ($\epsilon_r=1$) and dielectric media (constant $\epsilon_r$),\cite{Taflove}
\begin{equation}
\begin{cases}
a_{x1} = a_{y1}=1,\\
a_{x2} = -a_{y2}=\Delta t/(\epsilon_0\epsilon_r \Delta x),\label{eqn:d}\\
a_{x3} = a_{x4} = a_{x5} = a_{y3} = a_{y4} = a_{y5}=0,
\end{cases}
\end{equation}
in Drude media,
$\epsilon_m=\epsilon(\infty)-\omega_D^2/(\omega^2+i\Gamma_D\omega)$,\cite{Gray}
\begin{equation}
\begin{cases}
a_{x1} = a_{y1}=1,\\
a_{x2} = -a_{y2}=\Delta t/\left[\epsilon_0\epsilon(\infty) \Delta x\right],\\
a_{x3} = a_{y3} = -\Delta t/\left[\epsilon_0\epsilon(\infty)\right],\\
a_{x4} = a_{y4} = (1-\Gamma_D\Delta t)/(1+\Gamma_D\Delta t),\\
a_{x5} = a_{y5}= \epsilon_0\omega_D^2\Delta t/(1+\Gamma_D\Delta t),\label{eqn:e}
\end{cases}
\end{equation}
and in the PML region,
\begin{equation}
\begin{cases}
a_{x1} = \exp(-\sigma_y\Delta t/\epsilon_0),\\
a_{x2} = \left[1-\exp(-\sigma_y\Delta t/\epsilon_0)\right]/(\Delta x\sigma_y),\\
a_{x3} = a_{x4} = a_{x5} = 0;\\
a_{y1} = \exp(-\sigma_x\Delta t/\epsilon_0),\\
a_{y2} = -\left[1-\exp(-\sigma_x\Delta t/\epsilon_0)\right]/(\Delta x\sigma_x),\\
a_{y3} = a_{y4} = a_{y5} = 0.\label{eqn:f}
\end{cases}
\end{equation}
The coefficients in the $H$-field updating equations outside the PML regions are,
\begin{equation}
\begin{cases}
b_{x1} = b_{y1}=1,\\
b_{x2} = -b_{y2}=-\Delta t/(\mu_0\Delta x),\label{eqn:g}
\end{cases}
\end{equation}
whereas in the PML regions they read,
\begin{equation}
\begin{cases}
b_{x1} = \exp(-\sigma^*_x\Delta t/\mu_0),\\
b_{x2} = -\left[1-\exp(-\sigma^*_x\Delta t/\mu_0)\right]/(\Delta x\sigma^*_x),\\
b_{y1} = \exp(-\sigma^*_y\Delta t/\mu_0),\\
b_{y2} = \left[1-\exp(-\sigma^*_y\Delta t/\mu_0)\right]/(\Delta x\sigma^*_y).\label{eqn:h}
\end{cases}
\end{equation}
Here, we assume a polynomial grading of the PML parameters:\cite{Taflove}
$\sigma_{x,y}=\epsilon_0\sigma^*_{x,y}/\mu_0=\sigma_m(\rho / \delta)^m$, where $\sigma_m$ is the maximum conductance
in the PML, $\rho$ is the distance into the PML, and $\delta$ is the thickness
of the PML region. In this paper, we use a power $m=4$
and $\delta=20\Delta x$. $\sigma_m$ is optimized to give
a maximum reflection error on the order of $10^{-7}$.

The FDTD equations based on the UPML formulation read,\cite{Gedney}
\begin{eqnarray}
P_x\vert^{n+1}_{i+1/2,j}&=&P_x\vert^{n}_{i+1/2,j}
                           +\alpha_{x1}(H_z\vert^{n+1/2}_{i+1/2,j+1/2}-H_z\vert^{n+1/2}_{i+1/2,j-1/2}),\label{eqn:l}\\
D_x\vert^{n+1}_{i+1/2,j}&=&\alpha_{x2}D_x\vert^{n}_{i+1/2,j}+\alpha_{x3}D_x\vert^{n-1}_{i+1/2,j}
                           +\alpha_{x4}P_x\vert^{n+1}_{i+1/2,j}\notag\\
                           &&+\alpha_{x5}P_x\vert^{n}_{i+1/2,j}
                           +\alpha_{x6}P_x\vert^{n-1}_{i+1/2,j},\label{eqn:j}\\
E_x\vert^{n+1}_{i+1/2,j}&=&\alpha_{x7}E_x\vert^{n}_{i+1/2,j}
                           +\alpha_{x8}\alpha_{x9}D_x\vert^{n+1}_{i+1/2,j}
                           +\alpha_{x8}\alpha_{x10}D_x\vert^{n}_{i+1/2,j},\label{eqn:k}\\
P_y\vert^{n+1}_{i,j+1/2}&=&P_y\vert^{n}_{i,j+1/2}
                           +\alpha_{y1}(H_z\vert^{n+1/2}_{i+1/2,j+1/2}-H_z\vert^{n+1/2}_{i-1/2,j+1/2}),\\
D_y\vert^{n+1}_{i,j+1/2}&=&\alpha_{y2}D_y\vert^{n}_{i,j+1/2}+\alpha_{y3}D_y\vert^{n-1}_{i,j+1/2}
                           +\alpha_{y4}P_y\vert^{n+1}_{i,j+1/2}\notag\\
                           &&+\alpha_{y5}P_y\vert^{n}_{i,j+1/2}
                           +\alpha_{y6}P_y\vert^{n-1}_{i,j+1/2},\\
E_y\vert^{n+1}_{i,j+1/2}&=&\alpha_{y7}E_y\vert^{n}_{i,j+1/2}
                           +\alpha_{y8}\alpha_{y9}D_y\vert^{n+1}_{i,j+1/2}
                           +\alpha_{y8}\alpha_{y10}D_y\vert^{n}_{i,j+1/2},\label{eqn:k1}\\
B_z\vert^{n+3/2}_{i+1/2,j+1/2}&=&\beta_1 B_z\vert^{n+1/2}_{i+1/2,j+1/2}\notag\\
                              &&+\beta_2 (E_y\vert^{n+1}_{i+1,j+1/2}-E_y\vert^{n+1}_{i,j+1/2}
                                     -E_x\vert^{n+1}_{i+1/2,j}+E_x\vert^{n+1}_{i+1/2,j+1}),\label{eqn:i0}\\
H_z\vert^{n+3/2}_{i+1/2,j+1/2}&=&\beta_3 H_z\vert^{n+1/2}_{i+1/2,j+1/2}
                                 +\beta_4 (B_z\vert^{n+3/2}_{i+1/2,j+1/2}-B_z\vert^{n+1/2}_{i+1/2,j+1/2}).\label{eqn:i}
\end{eqnarray}

The coefficients in the $E$-field updating equations in all media are,
\begin{equation}
\alpha_{x1} = -\alpha_{y1}= \Delta t / \Delta x.
\end{equation}
In vacuum ($\epsilon_r=1$) and dielectric media (constant $\epsilon_r$),
\begin{equation}
\begin{cases}
\alpha_{x2}=\alpha_{x3}=\alpha_{x5}=\alpha_{x6}
=\alpha_{y2}=\alpha_{y3}=\alpha_{y5}=\alpha_{y6}=0,\\
\alpha_{x4}=\alpha_{y4}=1/\epsilon_r,
\end{cases}
\end{equation}
whereas in Drude media,
\begin{equation}
\begin{cases}
\alpha_{x2}=\alpha_{y2}=\frac{4\epsilon(\infty)}{2\epsilon(\infty)
                        +\epsilon(\infty)\Gamma_D\Delta t+\omega_D^2\Delta t^2},\\
\alpha_{x3}=\alpha_{y3}=\frac{-2\epsilon(\infty)
                        +\epsilon(\infty)\Gamma_D\Delta t-\omega_D^2\Delta t^2}{2\epsilon(\infty)
                        +\epsilon(\infty)\Gamma_D\Delta t+\omega_D^2\Delta t^2},\\
\alpha_{x4}=\alpha_{y4}=\frac{2+\Gamma_D\Delta t}{2\epsilon(\infty)
                        +\epsilon(\infty)\Gamma_D\Delta t+\omega_D^2\Delta t^2},\\
\alpha_{x5}=\alpha_{y5}=\frac{-4}{2\epsilon(\infty)
                        +\epsilon(\infty)\Gamma_D\Delta t+\omega_D^2\Delta t^2},\\
\alpha_{x6}=\alpha_{y6}=\frac{2-\Gamma_D\Delta t}{2\epsilon(\infty)
                        +\epsilon(\infty)\Gamma_D\Delta t+\omega_D^2\Delta t^2}.\\
\end{cases}
\end{equation}
outside the UPML regions,
\begin{equation}
\begin{cases}
\alpha_{x7}=\alpha_{x9}=\alpha_{x10}=\alpha_{x7}=\alpha_{x9}=\alpha_{x10}=1,\\
\alpha_{x8}=\alpha_{y8}=1/\epsilon_0,\\
\end{cases}
\end{equation}
and in the UPML regions,
\begin{equation}
\begin{cases}
\alpha_{x7}=\frac{2\epsilon_0\kappa_y-\sigma_y\Delta t}{2\epsilon_0\kappa_y+\sigma_y\Delta t},\\
\alpha_{x8}=\frac{1}{2\epsilon_0^2\kappa_y+\epsilon_0\sigma_y\Delta t},\\
\alpha_{x9}=\sigma_x\Delta t+2\epsilon_0\kappa_x,\\
\alpha_{x10}=\sigma_x\Delta t-2\epsilon_0\kappa_x,\\
\alpha_{y7}=\frac{2\epsilon_0\kappa_x-\sigma_x\Delta t}{2\epsilon_0\kappa_x+\sigma_x\Delta t},\\
\alpha_{y8}=\frac{1}{2\epsilon_0^2\kappa_x+\epsilon_0\sigma_x\Delta t},\\
\alpha_{y9}=\sigma_y\Delta t+2\epsilon_0\kappa_y,\\
\alpha_{x10}=\sigma_y\Delta t-2\epsilon_0\kappa_y.\label{eqn:j1}
\end{cases}
\end{equation}
The coefficients in the $H$-field updating equations outside the UPML region are,
\begin{equation}
\begin{cases}
\beta_1 = \beta_3 = 1,\\
\beta_2 = \Delta t / \Delta x, \\
\beta_4 = 1/\mu_0,
\end{cases}
\end{equation}
and in the UPML regions,
\begin{equation}
\begin{cases}
\beta_1 = \frac{2\epsilon_0\kappa_x-\sigma_x\Delta t}{2\epsilon_0\kappa_x+\sigma_x\Delta t},\\
\beta_2 = -\frac{2\epsilon_0 \Delta t}{(2\epsilon_0\kappa_x+\sigma_x\Delta t)\Delta x},\\
\beta_3 = \frac{2\epsilon_0\kappa_y-\sigma_y\Delta t}{2\epsilon_0\kappa_y+\sigma_y\Delta t},\\
\beta_4 = \frac{2\epsilon_0}{(2\epsilon_0\kappa_y+\sigma_y\Delta t)\mu_0}.\label{eqn:k2}
\end{cases}
\end{equation}
Here, we assume a polynomial grading of the PML parameters,\cite{Taflove}
$\sigma_{x,y}=\sigma_m(\rho / \delta)^m$ and $\kappa_{x,y}=1+(\kappa_m-1)(\rho / \delta)^m$, where
$\sigma_m$ and $\kappa_m$ denote the maxima of the UPML parameters
$\rho$ is distance into the PML, and $\delta$ is the thickness
of the PML. In this paper, we use power $m=4$, $\delta=20\Delta x$,
$\kappa_m=1$, and $\sigma_m$ is optimized to give
a maximum reflection error on the order of $10^{-7}$ for a simulation
region consisting of vacuum and on the order of $10^{-4}$ for a simulation
region consisting of Drude dispersive media.

\section{Systematic solution of one-dimensional wave propagation in the TM mode}
\label{Appx3}
In this appendix we provide a systematic solution for Eq. (\ref{eqn:eqn7}) when
$0<\sin(\theta)<1$ and
$\epsilon(\omega)$ consists of a linear superposition of Debye, Drude, and Lorentz
types of poles. In this case, we first rearrange  Eq. (\ref{eqn:eqn7}) as
\begin{equation}
\epsilon_{1r}\sin^2(\theta)H_z=\epsilon(\omega)(H_z-H^\prime_z),
\label{eqn:eqn34}
\end{equation}
where
\begin{equation}
\epsilon(\omega)=\epsilon(\infty)+\displaystyle\sum_{i}\epsilon_i(\omega),
\label{eqn:eqn35}
\end{equation}
and\cite{Taflove}
\begin{equation}
\epsilon_i(\omega) =
 \begin{cases}
 \epsilon_{DB}(\omega)=\frac{\Delta\epsilon_{DB}}{1-i\omega\tau_{DB}}, & \text{for a Debye pole} \\
 \epsilon_{DR}(\omega)=-\frac{\omega^2_{DR}}{\omega^2+i\Gamma_{DR}\omega}, & \text{for Drude pole pairs} \\
 \epsilon_{L}(\omega)=-\frac{\Delta\epsilon_{L}\omega^2_{L}}{\omega^2-\omega^2_{L}+i2\Gamma_{L}\omega},
                      & \text{for Lorentz pole pairs.} \\
 \end{cases}
 \label{eqn:eqn35a}
\end{equation}
We introduce auxiliary variables $K_i$ to
rewrite Eq. (\ref{eqn:eqn34}) as
a system of equations,
\begin{eqnarray}
\epsilon_{1r}\sin^2(\theta)H_z&=&K_0+\displaystyle\sum_{i}K_i,\label{eqn:eqn36}\\
K_0&=&\epsilon(\infty)(H_z-H^\prime_z),\label{eqn:eqn37}\\
K_i&=&\epsilon_i(\omega)(H_z-H^\prime_z).\label{eqn:eqn38}
\end{eqnarray}
Equations (\ref{eqn:eqn36}) and (\ref{eqn:eqn37}) correspond to
the set of FDTD equations,
\begin{gather}
\epsilon_{1r}\sin^2(\theta)H_z\vert^{n+3/2}-K_0\vert^{n+3/2}
-\displaystyle\sum_{i}K_i\vert^{n+3/2}=0,\label{eqn:eqn39}\\
\epsilon(\infty)H_z\vert^{n+3/2}-K_0\vert^{n+3/2}=\epsilon(\infty)H^{\prime}_z\vert^{n+3/2}.\label{eqn:eqn40}\\
\notag
\end{gather}
The translation of Eq. (\ref{eqn:eqn38}) into a set of FDTD equations depends
on the type of pole(s) considered [see Eq. (\ref{eqn:eqn35a})]. For a single Debye pole ($K_i=K_{DB}$),
\begin{eqnarray}
&&\Delta\epsilon_{DB}\Delta t H_z\vert^{n+3/2}-(2\tau_{DB}+\Delta t)K_{DB}\vert^{n+3/2}\notag\\
&&=(\Delta t - 2\tau_{DB})K_{DB}\vert^{n+1/2}\notag\\
&&+\Delta\epsilon_{DB}\Delta t(H^{\prime}_z\vert^{n+3/2}
+H^{\prime}_z\vert^{n+1/2}-H_z\vert^{n+1/2}).\label{eqn:eqn41}
\end{eqnarray}
For a Drude pole pair ($K_i=K_{DR}$),
\begin{eqnarray}
&&\omega^2_{DR}\Delta t^2 H_z\vert^{n+3/2}-(\Gamma_{DR}\Delta t+2)K_{DR}\vert^{n+3/2}\notag\\
&&=-4K_{DR}\vert^{n+1/2}+(2-\Gamma_{DR}\Delta t)K_{DR}\vert^{n-1/2}\notag\\
&&+\omega^2_{DR}\Delta t^2(H^{\prime}_z\vert^{n+3/2}+H^{\prime}_z\vert^{n-1/2}
-H_z\vert^{n-1/2}).\label{eqn:eqn42}
\end{eqnarray}
For a Lorentz pole pair ($K_i=K_{L}$),
\begin{eqnarray}
&&\Delta\epsilon_{L}\omega^2_{L}\Delta t^2 H_z\vert^{n+3/2}
-(\omega^2_{L}\Delta t^2+2\Gamma_{L}\Delta t+2)K_{L}\vert^{n+3/2}\notag\\
&&=-4K_{L}\vert^{n+1/2}+(\omega^2_{L}\Delta t^2-2\Gamma_{L}\Delta t+2)K_{L}\vert^{n-1/2}\notag\\
&&+\Delta\epsilon_{L}\omega^2_{L}\Delta t^2(H^{\prime}_z\vert^{n+3/2}+H^{\prime}_z\vert^{n-1/2}
-H_z\vert^{n-1/2}).\label{eqn:eqn43}
\end{eqnarray}
Equations (\ref{eqn:eqn39}) through (\ref{eqn:eqn43}) form a
linear system of equations for the unknowns $H_z\vert^{n+3/2}$
and $K_i\vert^{n+3/2}$ ($i=0,1,...$), which can be solved by existing numerical solvers for linear systems of equations.
The solution is then used to replace Eq. (\ref{eqn:eqn11}) to proceed the
1D wave propagation for the TM mode
Comparing to a direct Fourier transform
of Eq. (\ref{eqn:eqn5}), the above procedure only requires the storage
of the quantities at the previous two time instances and thus avoids the complexity of
numerical high-order derivatives with respect to time. This procedure
can be extended systematically to multi-poles in the material dispersiveness, although
it involves solving a linear system of equations.

\clearpage
\begin{center}
FIGURES
\end{center}

\clearpage

\clearpage

\clearpage

\clearpage

\clearpage

\clearpage

\clearpage

\clearpage

\begin{center}
TABLES
\end{center}

\begin{table}[htp]
\caption{\label{tab:tab1}Comparison of the magnitude of the reflection ($r$) and transmission ($t$) coefficients between the
analytical and numerical results for different incidence wavelengths ($\lambda$). Superscript $a$ denotes the analytical, and superscript $n$ denotes the numerical results.
The percentages in brackets denote the relative errors in the numerical results.
The mesh size is $\Delta x = 1$ nm and the Courant number is $S=0.3$.}
\begin{ruledtabular}
\begin{tabular}{ccccc}
 $\lambda$ (nm) & $|r^a|$ & $|r^n|$(error) & $|t^a|$ & $|t^n|$(error) \\ \hline
 $300$ & $0.2485$ & $0.2600(4.63\%)$ & $0.9686$ & $0.9658(0.29\%)$ \\
 $400$ & $0.1898$ & $0.1949(2.69\%)$ & $0.9818$ & $0.9809(0.09\%)$ \\
 $500$ & $0.1532$ & $0.1558(1.70\%)$ & $0.9882$ & $0.9878(0.04\%)$ \\
 $600$ & $0.1282$ & $0.1298(1.25\%)$ & $0.9912$ & $0.9916(0.04\%)$ \\
 $700$ & $0.6990$ & $0.6988(0.03\%)$ & $0.7152$ & $0.7154(0.03\%)$ \\
 $800$ & $0.7172$ & $0.7171(0.01\%)$ & $0.6969$ & $0.6969(<0.02\%)$\\
\end{tabular}
\end{ruledtabular}
\end{table}

\begin{table}[htp]
\caption{\label{tab:tab2}Comparison of the magnitude of the reflection ($r$) and transmission ($t$) coefficients between the
analytical and numerical results for different incidence  angles ($\theta$) and Courant numbers ($S$). Results
are obtained for a plane wave with $400$ nm wavelength
incident upon a $900$ nm thick dielectric slab ($\epsilon_r=11.7$). Superscript $a$ denotes the analytical, and superscript $n$ denotes the numerical results.
The percentages in brackets denote the relative errors in the numerical results.
The mesh size is $\Delta x = 1$ nm.}
\begin{ruledtabular}
\begin{tabular}{cccccc}
$\theta$ (degree) & $S$ & $|r^a|$ & $|r^n|$(error) & $|t^a|$ & $|t^n|$(error) \\ \hline
$0$ & $0.3$ & $0.8278$ & $0.8281(0.04\%)$ & $0.5610$ & $0.5539(1.27\%)$  \\
$10$ & $0.1$ & $0.8169$ & $0.8174(0.06\%)$ & $0.5767$ & $0.5760(0.12\%)$ \\
$20$ & $0.2$ & $0.7737$ & $0.7747(0.13\%)$ & $0.6335$ & $0.6323(0.21\%)$ \\
$30$ & $0.3$ & $0.6565$ & $0.6588(0.35\%)$ & $0.7543$ & $0.7526(0.23\%)$ \\
$40$ & $0.3$ & $0.3839$ & $0.3884(1.17\%)$ & $0.9234$ & $0.9216(0.19\%)$ \\
$50$ & $0.3$ & $0.0039$ & $0.0089(128\%)$ & $1.0000$ & $1.0000(<0.01\%)$ \\
$60$ & $0.3$ & $0.1981$ & $0.1952(1.46\%)$ & $0.9802$ & $0.9808(0.06\%)$ \\
$70$ & $0.2$ & $0.1152$ & $0.1142(0.87\%)$ & $0.9933$ & $0.9935(0.02\%)$ \\
$80$ & $0.1$ & $0.3395$ & $0.3367(0.82\%)$ & $0.9406$ & $0.9470(0.68\%)$ \\
\end{tabular}
\end{ruledtabular}
\end{table}

\begin{table}[htp]
\caption{\label{tab:tab3}As  in Table \ref{tab:tab1}
for oblique incidence upon
an $80$ nm thick silver slab. The Dielectric function of silver is described by the Drude model,
$\epsilon_m=\epsilon(\infty)-\omega_D^2/(\omega^2+i\Gamma_D\omega)$, with $\epsilon(\infty)=7.0246$,
$\omega_D=1.5713\times10^{16}$ rad/s, and $\Gamma_D=1.4003\times10^{14}$ rad/s.
The  mesh size is $\Delta x = 5$ nm and the  Courant number is $S=0.3$, except for the
$\lambda=350$ nm case, where $\Delta x = 2$ nm and the Courant number is $S=0.1$.}
\begin{ruledtabular}
\begin{tabular}{ccccc}
$\lambda$ (nm) & $|r^a|$ & $|r^n|$(error) & $|t^a|$ & $|t^n|$(error) \\ \hline
$350$ & $0.8908$ & $0.8908(<0.01\%)$ & $0.2295$ & $0.2295(<0.04\%)$ \\
$400$ & $0.9503$ & $0.9500(0.03\%)$ & $0.1231$ & $0.1234(0.24\%)$ \\
$450$ & $0.9672$ & $0.9672(<0.01\%)$ & $0.0760$ & $0.0761(0.13\%)$ \\
$500$ & $0.9743$ & $0.9742(0.01\%)$ & $0.0532$ & $0.0533(0.19\%)$ \\
\end{tabular}
\end{ruledtabular}
\end{table}

\begin{table}[htp]
\caption{\label{tab:tab4}As in Table \ref{tab:tab2}
for oblique incidence upon
an $80$ nm thick silver slab. The dielectric function of silver is described by the Drude model,
$\epsilon_m=\epsilon(\infty)-\omega_D^2/(\omega^2+i\Gamma_D\omega)$, with $\epsilon(\infty)=7.0246$,
$\omega_D=1.5713\times10^{16}$ rad/s, and $\Gamma_D=1.4003\times10^{14}$ rad/s.
The mesh size is $\Delta x = 5$ nm.}
\begin{ruledtabular}
\begin{tabular}{cccccc}
$\theta$ (degree) & $S$ & $|r^a|$ & $|r^n|$(error) & $|t^a|$ & $|t^n|$(error) \\ \hline
$0$ & $0.3$ & $0.9602$ & $0.9592(0.10\%)$ & $0.1202$ & $0.1201(0.08\%)$ \\
$10$ & $0.1$ & $0.9597$ & $0.9596(0.01\%)$ & $0.1204$ & $0.1204(<0.08\%)$ \\
$20$ & $0.2$ & $0.9580$ & $0.9581(0.01\%)$ & $0.1209$ & $0.1210(<0.08\%)$ \\
$30$ & $0.3$ & $0.9554$ & $0.9560(0.06\%)$ & $0.1218$ & $0.1220(0.16\%)$  \\
$40$ & $0.3$ & $0.9520$ & $0.9520(<0.01\%)$ & $0.1228$ & $0.1230(0.16\%)$  \\
$50$ & $0.3$ & $0.9488$ & $0.9498(0.11\%)$ & $0.1230$ & $0.1234(0.33\%)$  \\
$60$ & $0.3$ & $0.9478$ & $0.9475(0.03\%)$ & $0.1192$ & $0.1197(0.42\%)$  \\
$70$ & $0.2$ & $0.9532$ & $0.9531(0.01\%)$ & $0.1043$ & $0.1048(0.48\%)$  \\
$80$ & $0.1$ & $0.9711$ & $0.9627(0.87\%)$ & $0.0682$ & $0.0665(2.49\%)$  \\
\end{tabular}
\end{ruledtabular}
\end{table}


\begin{thebibliography}{99}
\bibitem{footnote1} We note that, when incoporated with the Convolutional
Perfectly Matched Layers absorbing boundary conditions, the
ADE approach can systematically treat a general dispersive medium
with a finite sum of Debye-, Lorentz- and Drude-type poles, while
the UPML approach with higher order poles becomes increasingly difficult
because of higher-order derivatives with respect to time. However,
for a single Drude pole considered in this paper, the UPML approach
is numerically tractable.
\bibitem{footnote3} IF $\epsilon(\omega)\leq\epsilon_{1r}\sin^2(\theta)$,
total reflection occurs, and the solution of Eqs. (\ref{eqn:eqn7})
and (\ref{eqn:eqn33})
becomes unstable. In the
examples provided in this paper, this situation is not allowed. Interested
readers are referred to Ref. \onlinecite{Capoglu} for a detailed discussion
of the FDTD formulation and solution in this case.
\end{thebibliography}
\end{document}